**Title:** Novel statistical ensemble analysis for simulating extrinsic noise-driven response in NF-κB signaling network

**Authors:** Jaewook Joo[1,]*, Steven J. Plimpton[2], Jean-Loup Faulon[3]

**Affiliations:** [1]Department of Physics and Astronomy, University of Tennessee, Knoxville, 37996, USA, [2]Scalable Algorithms Department, Sandia National Laboratories, Albuquerque, NM, 87185, USA; [3]Department of Biology, Every University, Every Cedex, France. *Corresponding author's email: jjoo1@utk.edu

**Abstract:**
Cellular responses in the single cells are known to be highly heterogeneous and individualistic due to the strong influence by extrinsic and intrinsic noise. Here, we are concerned about how to model the extrinsic noise-induced heterogeneous response in the single cells under the constraints of experimentally obtained population-averaged response, but without much detailed kinetic information. We propose a novel statistical ensemble scheme where extrinsic noise is regarded as fluctuations in the values of kinetic parameters and such fluctuations are modeled by randomly sampling the kinetic rate constants from a uniform distribution. We consider a large number of signaling system replicates, each of which has the same network topology, but a uniquely different set of kinetic rate constants. A protein dynamic response from each replicate should represent the dynamics in a single cell and the statistical ensemble average should be regarded as a population-level response averaged over a population of the cells. We devise an optimization algorithm to find the correct uniform distribution of the network parameters, which produces the correct statistical distribution of the response whose ensemble average and distribution agree well with the population-level experimental data and the experimentally observed heterogeneity. We apply this statistical ensemble analysis to a NF-κB signaling system and (1) predict the distributions of the heterogeneous NF-κB (either oscillatory or non-oscillatory) dynamic patterns and of the dynamic features (e.g., period), (2) predict that both the distribution and the statistical ensemble average of the NF-κB dynamic response depends sensitively on the dosage of stimulant, and lastly (3) demonstrate the sigmoidally shaped dose-response from the statistical ensemble average and the individual replicates.

**Key Words:** Statistical ensemble; extrinsic noise; NF-κB signal transduction

## I. INTRODUCTION

**Evidence of heterogeneous cellular response and its effect on cell fate decision:**
Technical advances in fluorescence imaging have sparkled and generated a surge of interest in cellular response at the single-cell level (Elowitz 2002; Blake 2003; Cohen et al 2008). This technique has already uncovered significant cell-to-cell variation in both gene expression (Elowiz 2002; Blake 2003; Raser and O'Shea 2005; Raj et al 2006) and protein dynamics (Cohen et al 2008; Lahav et al 2005). Recent theoretical and experimental work reveals that this cell-to-cell variability originates from both intrinsic (McAdams and Shapiro 1995; Arkin et al 1998; Thattai and van Oudenaarden 2001) and extrinsic noise (Swain 2002; Rosenfeld et al 2005). Noise critically affects cell-fate decision in developmental processes (Arkin et al 1998) and drug resistance (Cohen et al 2008). Moreover, response averaged over a population of the cells is oftentimes noticeably disparate from that in the single cells. The rational link between these two quantities needs to be established. Here, we propose a novel statistical method to unravel noise-induced discrepancy between single cell level and population level responses and to model the noise-driven heterogeneous cellular response at the single cells.

**Criticism:** The current modeling framework utilizes the dynamic, either deterministic or stochastic, models to unravel and predict the dynamic response of the biological networks. Most of the dynamic models contain many unknown kinetic rate constants, which need to be parameterized. A conventional parameterization scheme primarily focuses on identification of a single set of network parameters, which optimize the distance between the experimental data and the model prediction. As a result, the cellular response of the model is typically represented by a single dynamic pattern. This parameterization approach is a consequence of a strong influence by a deterministic viewpoint prevalent in biology: a biological response to a known stimulus should be robustly homogeneous and uniform and absolutely predictable. "Sloppy Cell", however, identifies a large number of the sets of the kinetic parameters of the biological network under the relaxed constraint that the model prediction is fitted to the experimental data within the experimental errors (Brown et al 2004). This is a clever fitting algorithm, but is still based on the same "homogeneous response" assumption and isn't designed to model the single cell behaviors. But, with the advancement and the availability of the single-cell level imaging techniques, the assumptions behind the above modeling methods are proven to be incorrect; the cellular response at the single cell level is heterogeneous and must be represented by a distribution of those heterogeneous responses. Therefore, the current modeling method needs to be modified to incorporate this single cell level property.

**Modeling of extrinsic noise and previous efforts:** In general, both intrinsic and extrinsic noises are the source of heterogeneous cellular responses. The intracellular signaling networks are stochastic generators because they typically consist of a small copy number of their constituents and are constantly under large fluctuations in the copy number of their constituents. Moreover, the signaling networks are exposed to time-varying environmental conditions and/or coupled to other unknown stochastic signaling networks. The former is called intrinsic noise because it is related only to a randomness of the reaction events. The effect of intrinsic noise on the cellular response is widely studied (Arkin et al 1998). The latter is extrinsic noise and its origins are unknown, but the few known origins are cell cycle (Rosenfeld et al 2005) and fluctuations in the copy number of transcriptional regulators in the upstream of the signaling network (Volfson et al 2006). Extrinsic noise affects all the constituents of the signaling network simultaneously and in a correlated manner. Several attempts were made to model extrinsic noise: to name a few, extrinsic noise is modeled as fluctuations in the network parameter values (Paulsson 2004) and the Gillespie's standard stochastic simulation algorithm (Gillespie 1997) is modified to take into account the time-varying kinetic rate constants (Swain's group 2008).

**Uncertainty in network parameter values:** Most of the biological dynamic models require the knowledge of a functional form and a kinetic rate constant of the associated reaction in a biological network. The lack of this critical kinetic information is one of the main drawbacks of the dynamic models and hampers their predictability. Thus, there is a great need to devise a modeling scheme, which doesn't require the precise kinetic information of a biological network. Extrinsic noise and the resulting heterogeneous response suggest that the kinetic rate constants should reflect uncertainty in their values, partly because of uncertainty in experimental measurements and partly because of their connection to the external fluctuations with unknown origins. Taking into account extrinsic noise, we can readily relax a strict constraint imposed to the parameterization of the kinetic parameters of the dynamic models.

**Novel statistical ensemble analysis:** We propose a statistical ensemble (SE) approach, which not only overcomes the lack of precise knowledge of network parameter values, but also provides a way to model extrinsic noise-driven heterogeneous cellular dynamic responses. We find a concept of the SE in the statistical physics useful in describing extrinsic noise and its effect on the resulting cellular responses. In our view, a biological system should be regarded as a complex system comprising a large number of components and elementary interactions among them and only the macroscopic state of such a complex bio-system is observable while its microscopic kinetic details are hidden and unknown to the observers. Taking extrinsic noise as fluctuations in kinetic parameter values (Paulsson Nature 2004), we propose a novel SE analysis. First, a population of the cells is modeled with a SE of a bio-system, i.e., a large number of the copies of identical system. Second, we assume that individual cells influenced by extrinsic noise operate and function in a biologically feasible range of kinetic parameter values. To simulate extrinsic noise, different values of kinetic parameters are sampled from a biologically feasible distribution and assigned to each system copy. Each copy corresponds to an individual single cell influenced by extrinsic noise. Each copy is unique in its kinetic parameter values and thus responds to the same stimulus diversely. The resulting dynamic response is no longer a single output but is represented by a distribution of heterogeneous responses from a SE of the system. We assign the same weight to each copy to calculate a macroscopic state and obtain the SE average of the system's response, which should be regarded as experimental data averaged over a population of the cells. The macroscopically observable response, the SE average, may take a simply behaving dynamic pattern. But, individual cell behavior can be very irregular, dissimilar, and diverse. The sum function (the SE average) annihilates any individualistic, heterogeneous, and non-smooth behavior. Thus, this SE analysis not only unravel the extrinsic noise-driven disparate behaviors between single cells and a population of the cells, but also can predict individual cell behaviors constrained by the experimental population level data. An important technical question is how to identify the biologically feasible and true distribution of extrinsic noise, which the kinetic parameters should be sampled from. Here we simplify the distribution to be uniform and our task to find the location and the interval size of this uniform distribution. We apply this statistical ensemble analysis to NF-κB signaling system.

**Application of statistical ensemble analysis to NF-κB signaling system:** NF-κB is a pleiotropic regulator of gene control and plays significant roles on various cellular functions such as differentiation of immune cells, development of lymphoid organs, and immune activation (Hoffmann et al., 2006; Verma and Stevenson, 2006). NF-κB shuttling between nucleus and cytoplasm is auto-regulated by the NF-κB signaling module, which consists of IκB (inhibitor κB), IKK (IκB kinase), and NF-κB. In the absence of stimulus, IκB forms a hetero-dimeric complex with NF-κB, preventing NF-κB from entering into the nucleus. Upon stimulation, the phosphorylated IKK catalyses the degradation of IκB from the IκB-NF-κB complex and frees up NF-κB whose nuclear localization initiates transcription of NF-κB target genes such as inflammatory cytokines (TNFα, IL-1, IL-6), chemotactic cytokines (MIP-1a), Th1 and Th2 response activation (IFN and IL-10), and lastly but most importantly negative regulators (IκBα, IκBβ, IκBε, and A20) which terminates the NF-κB signaling. Based upon the up-to-dated knowledge of NF-κB signaling, Hoffmann et al constructed a complex biochemical reaction network of NF-κB signaling consisting of IKK, NF-κB, and three IκB isoforms and transformed it into a set of ordinary differential equations with dozens of unknown kinetic parameters (Hoffmann et al., 2002). After identifying a single set of parameter values yielding the best fitting of the model to population level experimental data, they used their model to corroborate their argument about the role of each of three IκB isoforms: IκBα induces oscillatory shuttling of NF-κB while IκBβ and IκBε make the oscillation damped (Hoffmann et al., 2002). Lipniacki et al. computationally demonstrated that an additional negative regulator A20 has a definitive role as NF-κB signal terminator, assuming that A20 inactivates IKK phosphorylation (Lee et al., 2002; Lipniacki et al., 2004). Using the single cell fluorescence imaging, Nelson et al showed a remarkable phenomena at the level of single cells: some cells exhibits sustained oscillatory shuttling of NF-κB while others does non-oscillatory patterns, i.e., heterogeneous cellular response (Nelson et al. 2005; Lahav et al. 2004). Though NF-κB shuttling patterns are sustained-oscillatory at the single cells, their population level response is highly damped-oscillatory because averaging over a population of the cells masks oscillations at individual cells. However, Barken et al argued that, with the supportive experimental data, NF-κB dynamic responses at individual cells are highly synchronized and homogeneous (Barken et al., 2005). Hayot and Jayaprakash showed that intrinsic noise can unravel the mechanism behind the discrepancy between the oscillatory behaviors at single cells and the damped-oscillation at a population of the cells (Hayot and Japaprakash 2006).

**Summary:** We thoroughly investigate the effect of extrinsic noise on key protein dynamics in IKK-NF-κB-IκB-A20 signaling system. This signaling network is presented in Fig. 1 and consists of IKK, cytoplasmic and nuclear NF-κB, and two groups of negative regulators such as three isoforms of IκB and A20. Using the statistical ensemble analysis, we demonstrate that extrinsic noise, modeled as fluctuations in network kinetic parameters, derive the large deviation of the individual cell-level response away from their ensemble average. An ensemble of the NF-κB signaling system is generated by the Statistical Ensemble Generation method. This NF-κB network ensemble is stimulated by either large or small signal strength. In section II.1 we demonstrate that the statistical ensemble average of the individual profiles of each key biochemical species of the NF-κB signaling network is well fitted to experimental population level data for wild type and various mutant cases. In section II.2, we present a statistical analysis of the individual nuclear NF-κB profiles from the ensemble of a wild type and the mutants: the distribution of the dynamic characteristics such as dynamic patterns and dynamic features. In section II.3, we make a prediction about the dosage-dependent NF-κB response at the single cell levels: dosage-dependent statistical ensemble average behavior, dosage-dependent distribution of dynamic patterns, and that of dynamic features. Lastly, in section II.4, we show that both statistical ensemble average of dose-response curve and each individual curve are sigmoidally shaped.

## II. RESULTS

### 1. Fitting population level experimental data with their SE average of NF-κB signaling network

**A. Wild type case:** For the wild type case, we demonstrate that the SE scheme generates the significant cell-to-cell variability in protein dynamics, while successfully making the SE averages agree well with the population-level experimental data (EMSA or western blot or northern blot) of the key biochemical species as shown in Fig. 2. For the nuclear NF-κB profiles in Fig. 2 (a), the individual timings of the first peak are almost identical while the individual amplitudes of the first peak vary significantly with the deviation up to 100 % of its SE average. However, both the timings and amplitudes of the subsequent peaks exhibit a very significant cell-to-cell variability, consequently causing its SE average to take a damped oscillatory pattern with an outstanding peak followed by rapidly decaying amplitudes of subsequent peaks. Thus, the extrinsic noise produces "the masking effect of averaging over a population of asynchronous curves" just like the intrinsic noise does (Hayot et al JTB 2006).

**B. Fluctuations in IKK and IκB concentration as source of cell-to-cell variability in wild type case:** The large variation in the first peak amplitude of the nuclear NF-κB as shown in Fig. 2(a) originates from the IKK profile in Fig. 4 (b) where the individual curves of IKK concentration look a lot alike, yet with a great difference in their first peak amplitude which is laterally transferred to the large variation in the first valley of IκB isoforms as shown in Fig. 2 (b) and (c). Thus, the cell-to-cell variation in kinetic rate constants regulating the levels of both the pre-activated IKK (IKKn) and the activated of IKK (IKKa) is the source of the cell-to-cell variation in the first peak of nuclear NF-κB.

Likewise, the asynchronous behavior of the individual nuclear NF-κB profiles after two hours as shown in Fig. 2(a) originates from the cell-to-cell variability in various kinetic reactions: for example, it is cause by the variation in the second peak of the IκB isoforms as shown in Fig. 2 (b)-(d), which is caused by the variations in both the pre-stimulated level of IκB isoforms, by the variation in the first valley of IκB isoforms due to the cell-to-cell variability in IKK, and by the variation in the kinetic rate constants regulating the formation of NF-κB-IκB complex. We demonstrate that the dynamic characteristics of heterogeneous individual profiles of nuclear NF-κB is due to the extrinsic fluctuations in kinetic rate constants regulating the levels of IKK, IκBα, IκBβ, IκBε, NF-κB-IκB complex.

**C. IκB isoforms double knocked-out and A20 knocked-out mutants:** For the case of IκB isoforms double knocked-out and A20 knocked-out mutants, we set the synthesis rate of the IκB isoforms and A20 mRNA to zero, respectively. For the IκBβ and IκBε knocked-out mutant as shown in Fig. 3(a), the individual profiles of nuclear NF-κB are much more oscillatory (about a half of the curves are sustained oscillatory in Fig. 8) than those of wild case (only less than 10 % are sustained-oscillatory in Fig. 8). But, the SE average takes a damped oscillatory pattern and is a little more dynamical than that of the wild type case. This is mainly due to" the masking effect of averaging over a population of asynchronous curves". The peaks of the SE average correspond to the peaks from population level experimental data by Electro Mobility Shift Assay (EMSA) at 15 min, 2.5 hours, and 4 hours. For the IκBα and IκBε knocked-out and the IκBα and IκBβ knocked-out mutants as shown in Fig. 3(b) and 3(c), both SE averages of nuclear NF-κB take the similar looking "single-peaked" patterns as the population level EMSA measurements. For these two mutants, all the individual curves look alike to each other and each curve differ from the others only in its nuclear NF-κB level while the deviation of the individual curves from the SE average is as large as the 100 % of the SE average. For the A20 knocked-out mutant as shown in Fig. 4, both the SE averages of all the biochemical species take a single-peaked pattern just like the population level experimental data do. Though individual profiles are very alike, the deviation of the individual curves from its SE average is as large as 100 % of its SE average. In summary, for all the knocked-out mutants, both the individual profiles and their SE average of nuclear NF-κB take a uniform single-peaked pattern, but the variation in the nuclear NF-κB is very large among the individual profiles due to extrinsic noise.

**D. Dependence of the SE average of nuclear NF-κB on the heterogeneity measure:**
In Fig. 5 we show how the population level experimental measurements can place a strict restriction on the choice of the heterogeneity measure $\chi$, defined as the interval size of the uniform distribution from which each kinetic parameter is sampled. While fixing the kinetic rate constants at their reference values, we increase the heterogeneity measure and observe how heterogeneous the individual profiles of nuclear NF-κB become. The SE average of nuclear NF-κB becomes much less oscillatory for the higher value of heterogeneity measure as shown in Fig. 5. For a small value of heterogeneity measure in Fig. 5 (a) ($\chi$=10 %), we can clearly see that all individual curves are in phase with each other, making its SE average highly oscillatory too. For the higher values of heterogeneity measure in Fig. 5 (c) ($\chi$=50 %) and 5 (d) ($\chi$=70 %), a large fraction of individual curves are sustained oscillatory, but they are largely out of phase and asynchronous to each other and the resulting SE average is no longer oscillatory, but heavily under-damped oscillatory. Because the higher $\chi$ value assumes the larger sampling space, individual curves of nuclear NF-κB take the more heterogeneous dynamic patterns: some are sustained oscillatory while others are single-peaked. Thus, if the population level experimental measurements exhibit the oscillatory response, this data can place a strict restriction on the choice of the heterogeneity measure.

**2. Prediction of the SE distribution of the dynamic patterns and the dynamic features for both the wild type and the mutants**

**A. Distribution of the dynamic features:** In Fig. 6 we demonstrate that the SE analysis can capture the distribution of the dynamic features (see methods) of the individual cellular responses and its variation upon the knocked-out of the genes. It is also shown that there is a significant amount of overlap between the distribution of the wild type and that of the other mutants. This implies that if we rely on the conventional parameterization scheme which identifies a single set of parameters and present a single representative response, we can easily fall into a deceptively incorrect conclusion about the effect of gene knocked-out on the cellular response. To avoid such a pitfall, we represent the single cell level dynamic response with the distribution of their dynamic features and observe any significant change in the distribution when genes are knocked out. For Fig. 6 (a), the distributions of the First Maximum are invariant between the mutants and the wild type. This dynamic feature shouldn't be chosen as an indicator of the physiological defects due to gene knocked-out. For Fig. 6 (b), the A20 knocked-out mutant increases the average value of the First Translocation Time while the I$\kappa$B$\beta$ and I$\kappa$B$\epsilon$ knocked-out mutant decreases it. But, the wild type and two other mutants share the similar distribution. In Fig. 6 (c), only the wild type and the I$\kappa$B$\beta$ and I$\kappa$B$\epsilon$ knocked-out mutant have their averages of the First Period at about 2.2 and 2 hours respectively, and the other mutants have too broadly distributed First Period to define their averages. In Fig 6 (d), the ratio of the First Minimum to the First Maximum indirectly measures the spikiness of the oscillations defined as the deep valley between two adjacent maxima: The smaller this ratio gets, the spikier the temporal profile becomes. Only the wild type and the I$\kappa$B$\beta$ and I$\kappa$B$\epsilon$ knocked-out mutant exhibit the spiky responses. In Fig 7 (f), the ratio of the Steady State to the First Maximum provides useful information about the relative magnitude and strength of the negative regulators of I$\kappa$B isoforms and A20. Remembering that the distributions of the First Maximum are invariant between the wild type and the mutants, we conclude that the smaller steady state level of nuclear NF-$\kappa$B means the stronger negative feedback. We list the mutants in order of the steady state level: A20 knocked-out mutant < I$\kappa$B$\alpha$ and I$\kappa$B$\epsilon$ knocked-out mutant < I$\kappa$B$\alpha$ and I$\kappa$B$\beta$ knocked-out mutant < I$\kappa$B$\beta$ and I$\kappa$B$\epsilon$ knocked-out mutant < wild type. Now the order of the strength of the negative regulators can be inferred from the above ordered list: A20 > I$\kappa$B$\alpha$ > I$\kappa$B$\epsilon$ > I$\kappa$B$\beta$. This order is determined by our choice of the kinetic rate constants.

**B. Distribution of the dynamic patterns:** The individual time-series of the nuclear NF-$\kappa$B concentration can be classified into one of four dynamic patterns as shown in Fig. 7: damped oscillation, sustained oscillation, single peaked pattern, and hyperbolic pattern.

The underlying mechanism for each dynamic pattern is rather simple: The hyperbolic (or over-damped) pattern originates from the strong strength of the negative feedback loops. while the single-peaked pattern comes from the weak negative feedback loops. Oscillatory pattern arises from a moderate strength of negative feedback loops. But, it remains open to correlate each dynamic pattern with a cellular physiological condition. For the cases of the wild type and the mutants, we generate and stimulate the SE of the NF-κB network with the same signal strength (TR=1), respectively. Then we classify a thousand individual temporal profiles into one of four dynamic patterns. The distributions of the dynamic patterns are represented by the bar graphs in Fig. 7 and show that the SE of the wild type or the mutant takes at least two different dynamic patterns under the same stimulation. For wild type case, the distribution of the nuclear NF-κB profiles is very skewed to a single pattern, the damped-oscillatory pattern, while a less than 10 % of the profiles are sustained-oscillatory. This indicates that for the wild type case the damped oscillation is the most probable pattern and is robust against the perturbation of the network parameter values. For the A20 knocked-out mutant, both single-peaked and damped-oscillatory patterns are almost equally probable. But, those damped oscillatory profiles are very much like a single-peaked pattern. In other words, for this mutant, the damped-oscillation occurs at a particular region of the parameter space where the strength of the negative regulators is not strong enough to induce the oscillatory pattern. For IκBβ and IκBε knocked-out mutant, sustained-oscillatory and damped-oscillatory patterns are equally probable dynamic response. The damped-oscillatory patterns in this mutant are very different from those in the A20 knocked-out mutant and are similar to a sustained-oscillation. The fraction of the sustained oscillation (about 50 %) dramatically increases in comparison to the wild type case (less than 10 %). For both IκBα and IκBβ knocked-out and IκBα and IκBε knocked-out mutants, their respective distributions are similar to that of A20 knocked-out mutant in Fig. 7. As evidently shown in Fig. 3 (b) and 3 (c) and Fig. 4 (c), both the individual profiles and the statistical ensemble average of the nuclear NF-κB concentration for the A20 knocked-out, the IκBα and IκBβ knocked-out, and the IκBα and IκBε knocked-out mutants share the similar single peaked pattern. In summary, there are two distinctive groups exhibiting two respective dynamic patterns of nuclear NF-κB profiles: the first group, the wild type and the IκBβ and IκBε knocked-out mutant, show the highly oscillatory pattern and the second group, the A20 knocked-out, the IκBα and IκBβ knocked-out and the IκBα and IκBε knocked-out mutants, shows the (non-oscillatory) single-peaked pattern.

### 3. SE analysis of dosage-dependent NF-κB behavior

**A. Dosage-dependent SE average and individual profiles of nuclear NF-κB concentration:** The SE of NF-κB network is stimulated with either the large (TR=1) or the small (TR=0.01) signal strength, respectively. The dosage-dependent behavior of both the SE average and the individual temporal profiles of the biochemical species are presented in Fig. 8. The small dosage induces either monotone-increasing (hyperbolic) or single peaked IKK individual profiles as shown in Fig. 8 (b) while the large dosage make them take the single-peaked pattern possessing a prominent peak as shown in Fig. 8 (g). This dosage-dependent IKK profiles are directly transferred to the cytoplasmic IκBα profiles. For the small dosage, the hyperbolic IKK profiles induce the exponentially decaying IκBα profiles as shown in Fig. 8 (c). On the contrary, for the large dosage, the noticeable peaks of the IKK profiles suppress the cytoplasmic IκBα levels and, when the peaks of the IKK profiles drop down to their steady state levels, the cytoplasmic IκBα levels recover back to their equilibrium levels in a damped oscillatory manner as shown in Fig. 8 (h). The nuclear NF-κB profiles follow the IκBα profiles but in an exactly opposite way: the cytoplasmic IκBα sequesters NF-κB in the cytoplasm, inhibiting nuclear localization of NF-κB. Since the mRNA synthesis rate of the NF-κB target genes is assumed to be linearly dependent on the nuclear NF-κB concentration, the profiles of both A20 mRNA and IκBα mRNA follow the nuclear NF-κB profile after half an hour time lag.

**B. Dosage-dependent distribution of the dynamic features:** After stimulating the SE of the NF-κB system with the large (TR=1) or the small (TR=0.01) dosage, we obtain the distribution of each dynamic feature from a thousand nuclear NF-κB profiles as shown in Fig. 9. In Fig. 9 (a) and (c), both the First Maximum and the First Period share the similar dosage-dependent behavior: the smaller dosage induces the distribution mode located at the smaller First Maximum or the smaller First Period while the full half maximum width of the distribution is independent of the dosage. But, for both the First Translocation Time and the Ratio of the First Minimum to the First Maximum, the dosage-dependent behavior is opposite to the previous case as shown in Fig. 9 (b) and 9 (d): the larger dosage induces the distribution peaked at the earlier First Transition Time or at the smaller First Minimum level. Moreover, the larger dosage makes their distributions the more narrowly distributed. This indicates that the larger dosage induces the earlier and spikier response and the smaller dosage induces the more heterogeneous First Maximum and First Minimum levels of nuclear NF-κB. Lastly, both the ratios of the Second Maximum to the First Maximum and of the Steady State to the First Maximum share the similar dosage-dependent behavior as shown in Fig. 9 (e) and 9 (f): the smaller dosage induces the mode of the distribution at the larger values, i.e., closer to one. In other words, when stimulated by the smaller dosage, the levels of the First Maximum, of the subsequent maxima, and of the Steady State are the same, i.e., NF-κB profiles take either monotone-increasing pattern or single-peaked pattern with low peak amplitude. In addition, the full half maximum width of the distribution is unaffected by the change of the dosage.

**C. Dosage-dependent distribution of the dynamic patterns:** As shown in Fig. 10, when stimulated by the small (TR=0.01) dosage, 80 % of the nuclear NF-κB profiles are damped-oscillatory whereas only 20 % of them are single-peaked. But, those damped oscillatory patterns are a lot like a single-peaked pattern. The distribution induced by the large dosage (TR=1) corresponds to that of the wild type case in Fig. 7. We note that the distribution of the dynamic patterns, the SE ensemble average, and individual profiles of nuclear NF-κB concentration upon the small (TR=0.01) dosage stimulation in Figs. 8 and 10 are very similar to those from the IκBα and IκBε knocked-out mutant upon the large (TR=1) dosage stimulation in Figs. 3 and 7. When the heterogeneity measure χ increases from its present value χ=30 % to χ=70 %, the small dosage stimulation generates more heterogeneous dynamic patterns, i.e., more equally distributed dynamic patterns of the nuclear NF-κB.

**4. Sigmoidally shaped SE average of the dose-response curves**

We numerically investigate the distribution of the dose-response curves from the SE of the NF-κB system. We generate the SE of the NF-κB system with only 50 replicates because of high computational cost to get one dose-response curve. The SE of NF-κB signaling system is stimulated with a persistent signal strength for 30 hours duration and the average (quasi-) steady state level of nuclear NF-κB concentration between 20 and 30 hours after stimulation is measured. To test for the presence of the hysteresis effect, we compute the dose-response curve twice, i.e., increase the signal strength from *TR=0.1* to *TR=0* in a step-like manner and then decrease it from *TR=0.1* to *TR=0*. For each replicate in the SE, both forward and backward dose-response curves look exactly the same and this indicates the absence of hysteresis effect in the NF-κB signaling system and take a sigmoidal shape as shown in Fig. 11. The steady state nuclear NF-κB levels change dramatically at the inflection points of the sigmoidal curves in Fig. 11. For the signal strength smaller than the inflection point of each curve, the stationary nuclear NF-κB level is very low while for the signal just greater than the inflection point, the stationary level quickly reaches a plateau. Lastly, the SE average of those individual dose-response curves is sigmoidally shaped.

**III. DICUSSION**

**New predictions:** In this paper, we devise a novel statistical method to mimic the protein dynamics in a population of the cells under the influence of extrinsic noise. We demonstrate, after making the SE average to match with a population level experimental data, that the SE of the signaling system produces several experimentally observable distributions of the dynamic characteristics of the protein profiles. The main predictions are listed as follows: (a) under the same experimental condition, the nuclear NF-κB profiles are expected to take the heterogeneous dynamic patterns at the single cell level, (b) the larger dosage induces the more oscillatory dynamic patterns of the nuclear NF-κB while the smaller dosage does the single-peaked patterns, (c) the larger (smaller) dosage makes the First translocation time narrowly-distributed (broadly-distributed) and the peak of its distribution shifted to the earlier (later) time, and (d) the shape of the dose-response curves both at the single cell level and the population level is sigmoidal. Most of our predictions, e.g., (a) – (c), have been verified by our colleagues (reference). We hope to elicit more experimental efforts to verify the above predictive dynamic behavior in the single cells.

**Novel viewpoint of statistical ensemble analysis:** We like to emphasize that the novelty of our SE analysis lies not on the technical part, but on the new viewpoint on the cellular response. All the previous papers focus mainly on the analysis of the dynamic behavior of the model dynamic systems in a restricted parameter space by the usage of the Monte Carlo sampling of the kinetic parameter values and the classification and/or sensitivity analysis of the resulting dynamic response (reference; Joo et al 2007). But, we start with a totally different viewpoint that the protein dynamics of the individual cells are intrinsically heterogeneous because they are exposed mainly to extrinsic noise and likely to have large fluctuations in the kinetic parameters controlling the abundance of the biochemical species in the cells.

**Sigmoidally shaped dose-response curve:** The sigmoidal shape of the dose-response curve reveals two important properties of the NF-κB signaling: the switching behavior and the monostability (no hysteresis). The inflection points of the individual sigmoidal curves can play the role of the activation threshold of the NF-κB signaling pathway. As shown in Fig. 11, the NF-κB response is so little to the signal strength below the threshold while its response significantly increases for the signal strength just above the threshold. Knowing that some of the NF-κB target genes are the inflammatory cytokines and overly expressed inflammatory response is adversary to the host, we can speculate that the NF-κB signaling network employs this sigmoidal dose-response curve to down-regulate the excessive inflammatory response, i.e., turn it on only if the danger level is significantly high, otherwise shut it down. However, the amplitude and the timing of the first peak of the inflammatory cytokines such as TNFα are known to be critical to elicit the timely and effective immune response (Mann et al 2002). In this case, we'd better to measure the dosage-dependent transient dynamic response of NF-κB target genes and investigate the shape of the dose-(transient dynamic) response. Lastly, TNFα autocrine signaling forms the (+) feedback loop in the NF-κB signaling network and can induce the bistability, which may modify our results on the monostability.

**True distribution of extrinsic noise:** Our statistical analysis of the protein dynamics depends on how biologically and realistically the computationally generated SE of the NF-κB system represents a population of the individual cells. This question is equivalent to what is the true distribution of extrinsic noise, i.e., the distribution according to which the kinetic parameters should be sampled. In this paper, we simplify this problem greatly by assuming that this distribution is a uniform one. We devise a heuristic fitting algorithm to find the interval of the uniform distribution of the kinetic parameters by minimizing the discrepancy between the SE analysis and the population level experimental data. But, to make the sampling more biological, it should be taken into account that the distribution of extrinsic noise can change in time periodically as the cells undergo cell cycle. Moreover, though we assume no correlation between any pair of the kinetic parameter values, the parameters may be statistically dependent simply because the cellular energy resource must be limited: e.g., as one kinetic process gets accelerated, then the others should decrease in order to balance the cellular energy consumption. To simulate the protein dynamics in the real single cells, first, it is highly required to devise the single cell experimental techniques from which we can infer this true distribution of extrinsic noise. Second, the optimization problem needs to be solved rigorously: to find the distribution of the kinetic parameters of the dynamic network which minimizes the difference between the SE average and the population-level experimental measurements and simultaneously reproduce the experimentally observed heterogeneous protein dynamics in the single cells.

**5. Classification:** Our choice of only four dynamic patterns of nuclear NF-κB profiles greatly simplifies the statistical analysis and helps us observe clearly the change of the distribution upon gene knocked-out perturbations. Our choice is based on the dynamical and mathematical characteristics of the protein profile. However, it is possible that this simplification neglects the other biologically important details of the nuclear NF-κB profiles. If we go by a different choice of the dynamic patterns, e.g., classification by period or by steady state level, it may change the distribution of the dynamic patterns and its change upon perturbations.

## IV. METHODS

**Six dynamic features of nuclear NF-κB profile:** We define six dynamic features to represent the "unique" characteristics of temporal profiles of nuclear NF-κB concentration. The first translocation time depicts the time when the first peak occurs; the first period measures the time between the first two peaks; the first and the second maxima define the peak amplitudes of the first and the second peaks; the first minimum means the amplitude of the first valley; the steady state refers to the amplitude level at sufficiently long time. Making the first maximum level the reference, we present a scaled ratio, i.e., the level of the first minimum, the second maximum, and the steady state normalized by the first maximum. The distributions of the dynamical features are presented in Figs. 6 and 9.

**Generation of the SE of NF-κB signaling network:** Each kinetic rate constant listed in Table 1 is randomly sampled from an interval ($x_0(1-\chi), x_0(1+\chi)$) where $x_0$ is the reference value and $\chi$ is a heterogeneity measure. To enhance the efficacy of the sampling in the high dimensional space, we employ the Latin Hypercubic sampling method discussed in methods section. For this paper, we set $\chi = 0.3$. To generate the SE consisting of $N$ replicates, we simply make $N$ sets of randomly sampled kinetic parameters.

**Algorithm to fit the SE average to population level experimental data:** We will not attempt to fit the SE average to the entire time-series data. To try to do so results in notorious over-fitting: much more parameters to fit much less data. Moreover, biology data are rather qualitative than quantitative. Also, we learn that there are only a handful number of the kinetic parameters in the NF-κB signaling network whose variation significantly affects the temporal profile of the nuclear NF-κB concentration (Joo et al 2007). Our fitting algorithm is to fit the dynamic features of the SE average to those of the experimental time-series data. Based on our previous studies (Joo et al 2007), we choose two kinetic parameters most highly correlated with each dynamic feature and vary them until the dynamic features between the SE average and the experimental time-series data are matched. Here are the actual steps that we take: For step 1, use an educated guess for kinetic parameters and set the heterogeneity measure to $\chi = 0.9$. For step 2, generate the SE and the resulting protein dynamic profiles and calculate the deviation of the six dynamic features of the SE average from the target dynamic features. For step 3, identify the most deviated dynamic feature and modify two kinetic parameters associated with that dynamic feature. For step 4, repeat the steps one through three until the dynamic features get close to the target values. For step 5, when the good fitting is not achievable with the pre-set value of $\chi$, we decrease it in a step-like manner. All the Figures 2 through 5 are obtained through the above fitting algorithm.

**Numerical simulation of the NF-κB signaling network:** A system of ordinary differential equations is derived from the NF-κB signaling network in Fig. 1 and the kinetic parameters in Table 1. Using the Runge-Kutta 4$^{th}$ order, we numerically solve the dynamic model with the sampled kinetic parameters and with the initial conditions: the zero concentrations of all the other biochemical species and a sampled total concentration of cytoplasmic NF-κB. Before stimulating the system (*TR=0*), the dynamic system runs for 33 hours until its constituents reach their equilibrium values. Then, we simulate the persistent stimulation by turning on the reaction, IKKn → IKKa with a rate $TR \cdot K1$, i.e., by assigning a non-zero value to *TR,* where *TR* stands for the dosage of a stimulant.

**Latin Hypercube sampling (LHS):** LHS is a constrained Monte Carlo sampling scheme. The Monte Carlo sampling scheme is a conventional approach and a common choice for the uncertainty assessment of a computational model. By sampling repeatedly from the assumed joint probability function of the input variables and evaluating the response for each sample, the distribution of the response of the computer model can be estimated. This approach yields reasonable estimates for the distribution of the response if the number of samples is quite large. However, since a large sample size requires a large number of computations from the computer model (a potentially very large computational expense), an alternative approach, Latin Hypercube sampling, can be used. LHS yields more precise estimates with a smaller number of samples, and is designed to address the above concern (Swiler et al 2004). Suppose that the computer model has K kinetic rate variables and we want N samples. LHS selects N different values from each of K kinetic rate variables such that the range of each variable is divided into N non-overlapping intervals on the basis of equal probability. One value from each interval is selected at random with respect to the assumed probability density in the interval. The N values thus obtained for the first kinetic rate variable are paired in a random manner (equally likely combinations) with the N values of the second kinetic rate variable. These N pairs are combined in a random manner with the N values of the third kinetic rate variable to form N triplets, and so on, until N K-tuplets are formed. These N K-tuplets are the same as the N K-dimensional input vectors where the ith input vector contains specific values of each of the K kinetic rate variables to be used on the ith run of the computer model (Swiler et al 2004).

Table I. Biochemical reactions & associated reaction rates in our computational model of the NF-κB signaling network. The reaction rates labeled with [1] are from Ref. [32], those labeled [2] are from Ref. [26], those labeled [3] use an average value between those in Ref. [26] & Ref. [32]. Column *I* is the kinetic parameter, *II* is its units, *III* is the published nominal value, *IV* is the reference, and *V* is our choice of kinetic value after fitting. The units for [a] are $\mu M^{-1} s^{-1}$, for [b] are $s^{-1}$, for [c] are $\mu M\ s^{-1}$, and for [d] are $\mu M$.

| Reactions | I | II | III | IV | V |
|---|---|---|---|---|---|
| IKKa + IκBa → IKKa_IκBα | Aα | [a] | 0.2 | [1] | 0.1813 |
| IKKa + IκBb → IKKa_IκBβ | Aβ | [a] | 0.05 | [3] | 0.02997 |
| IKKa + IκBe → IKKa_IκBε | Aε | [a] | 0.05 | [3] | 0.04244 |
| IKKa+IkBα-NF-κB → IKKa-IκBα-NF-κB | Bα | [a] | 1 | [1] | 1.024 |
| IKKa+IkBβ-NF-κB → IKKa-IκBβ-NF-κB | Bβ | [a] | 0.25 | [3] | 0.3683 |
| IKKa+IkBε-NF-κB → IKKa-IκBε_NFkB | Bε | [a] | 0.25 | [3] | 0.42 |
| NF-κBn → NF-κBn + A20t | C1 | [b] | 0.0000005 | [1] | 0.000000506 |
| 0 → A20t | C2 | [c] | 0 | [1] | 0 |
| A20t → 0 | C3 | [b] | 0.0004 | [1] | 0.0002438 |
| A20t → A20t + A20 | C4 | [b] | 0.5 | [1] | 0.5807 |
| A20 → 0 | C5 | [b] | 0.0003 | [1] | 0.0003769 |
| IKKa-IκBα → IKKa + IκBα | Dα | [b] | 0.00125 | [2] | 0.002046 |
| IKKa-IκBβ → IKKa + IκBβ | Dβ | [b] | 0.00175 | [2] | 0.0005609 |
| IKKa-IκBε → IKKa + IκBε | Dε | [b] | 0.00175 | [2] | 0.002142 |
| IKKa-IκBα-NF-κB → IKKa + IκBα-NF-κB | Dα | [b] | 0.00125 | [2] | 0.002046 |
| IKKa-IkBβ-NF-κB → IKKa + IκBβ-NF-κB | Dβ | [b] | 0.00175 | [2] | 0.000561 |
| IKKa-IκBε-NF-κB → IKKa + IκBε-NF-κB | Dε | [b] | 0.00175 | [2] | 0.002142 |
| IKKa-IκBα-NF-κB → IKKa-IκBα + NF-κB | Eα | [b] | 0.000001 | [2] | 0.00000144 |
| IKKa-IκBβ-NF-κB → IKKa-IκBβ + NF-κB | Eβ | [b] | 0.000001 | [2] | 0.00000124 |
| IKKa-IκBε-NF-κB → IKKa-IκBε + NF-κB | Eε | [b] | 0.000001 | [2] | 0.00000064 |
| IKKa-IκBα + NF-κB → IKKa-IκBα-NF-κB | Fα | [a] | 0.5 | [2] | 0.3789 |
| IKKa-IκBβ + NF-κB → IKKa-IκBβ-NF-κB | Fβ | [a] | 0.5 | [2] | 0.2135 |
| IKKa-IκBε + NF-κB → IKKa-IκBε-NF-κB | Fε | [a] | 0.5 | [2] | 0.3528 |
| IκBα-NF-κB → NF-κB + IκBα | Gα | [b] | 0.000001 | [2] | 0.00000064 |
| IκBβ-NF-κB → NF-κB + IκBβ | Gβ | [b] | 0.000001 | [2] | 0.00000044 |
| IκBε-NF-κB → NF-κB + IκBε | Gε | [b] | 0.000001 | [2] | 0.00000069 |
| IκBαn-NF-κBn → NF-κBn + IκBαn | Gα | [b] | 0.000001 | [2] | 0.00000064 |
| IκBβn-NF-κBn → NF-κBn + IκBβn | Gβ | [b] | 0.000001 | [2] | 0.00000044 |
| IκBεn-NF-κBn → NF-κBn + IκBεn | Gε | [b] | 0.000001 | [2] | 0.00000069 |
| IκBα + NF-κB → IκBα-NF-κB | Hα | [a] | 0.5 | [2] | 0.4593 |
| IκBβ + NF-κB → IκBβ-NF-κB | Hβ | [a] | 0.5 | [2] | 0.7753 |
| IκBε + NF-κB → IκBε-NF-κB | Hε | [a] | 0.5 | [2] | 0.2895 |
| IκBαn + NF-κBn → IκBαn-NF-κBn | Hα | [a] | 0.5 | [2] | 0.4593 |
| IκBβn + NF-κBn → IκBβn-NF-κBn | Hβ | [a] | 0.5 | [2] | 0.7753 |
| IκBεn + NF-κBn → IκBεn-NF-κBn | Hε | [a] | 0.5 | [2] | 0.2895 |
| NF-κB → NF-κBn | I1 | [b] | 0.0025 | [1] | 0.003037 |
| NF-κBn → NF-κB | K01 | [b] | 0.00005 | [3] | 0.00005537 |
| IKKn → IKKa | K1 | [b] | 0.0025 | [1] | 0.003273 |
| A20 +IKKa → A20 + IKKi | K2 | [a] | 0.1 | [1] | 0.07075 |
| IKKa → IKKi | K3 | [b] | 0.0015 | [1] | 0.00202 |
| 0 → IKKn | Kprod | [c] | 0.000025 | [1] | 0.000009752 |
| IKKn, IKKa, or IKKi → 0 | Kdeg | [b] | 0.000125 | [1] | 0.0001561 |
| Volume ratio of cytoplasm to nucleus | Kv | 1 | 5 | [1] | Variable |
| IκBαn-NF-κBn → IκBα-NF-κB | Lα | [b] | 0.01 | [1] | 0.013979 |
| IκBβn-NF-κBn → IκBβ-NF-κB | Lβ | [b] | 0.005 | [3] | 0.001567 |
| IκBεn-NF-κBn → IκBε-NF-κB | Lε | [b] | 0.005 | [3] | 0.006583 |
| IκBα-NF-κB → NF-κB | Mα | [b] | 0.000025 | [1] | 0.00002837 |
| IκBβ-NF-κB → NF-κB | Mβ | [b] | 0.000025 | [3] | 0.00003609 |
| IκBε-NF-κB → NF-κB | Mε | [b] | 0.000025 | [3] | 0.00000866 |
| IKKa-IκBα-NF-κB → IKKa + NF-κB | Pα | [b] | 0.1 | [1] | 0.12928 |
| IKKa-IκBβ-NF-κB → IKKa + NF-κB | Pβ | [b] | 0.05 | [3] | 0.06454 |
| IKKa-IκBε-NF-κB → IKKa + NF-κB | Pε | [b] | 0.05 | [3] | 0.08434 |
| IκBαn → IκBα | Qα | [b] | 0.0005 | [1] | 0.0005123 |
| IκBβn → IκBβ | Qβ | [b] | 0.0005 | [3] | 0.0007398 |
| IkBεn → IkBε | Qε | [b] | 0.0005 | [3] | 0.0002184 |
| IKKa-IκBα → IKKa | Rα | [b] | 0.1 | [1] | 0.123 |
| IKKa-IκBβ → IKKa | Rβ | [b] | 0.1 | [3] | 0.03837 |
| IKKa-IκBε → IKKa | Rε | [b] | 0.1 | [3] | 0.1571 |
| IκBαn-NF-κBn → NF-κBn | Sα | [b] | 0.000001 | [2] | 0.00000037 |
| IκBβn-NF-κBn → NF-κBn | Sβ | [b] | 0.000001 | [2] | 0.000001131 |
| IκBεn-NF-κBn → NF-κBn | Sε | [b] | 0.000001 | [2] | 0.000001037 |
| NF-κBn → NF-κBn + IκBαt | Uα | [b] | 0.0000005 | [1] | 0.000000279 |
| NF-κBn → NF-κBn + IκBβt | Uβ | [b] | 0 | [2] | 0 |
| NF-κBn → NF-κBn + IκBεt | Uε | [b] | 0.00000005 | [3] | 0.000000059 |
| IκBα → IκBαn | Vα | [b] | 0.001 | [1] | 0.0009786 |
| IκBβ → IκBβn | Vβ | [b] | 0.001 | [3] | 0.0004871 |
| IkBε → IkBεn | Vε | [b] | 0.001 | [3] | 0.00147 |
| IκBα, IκBαn → 0 | Wα | [b] | 0.0001 | [1] | 0.000132 |
| IκBβ, IκBβn → 0 | Wβ | [b] | 0.0001 | [3] | 0.000133 |
| IκBε, IκBεn → 0 | Wε | [b] | 0.0001 | [3] | 0.000042 |
| IκBαt → IkBαt + IkBα | Xα | [b] | 0.5 | [1] | 0.4552 |
| IκBβt → IκBαt + IκBβ | Xβ | [b] | 0.5 | [3] | 0.3828 |
| IκBεt → IκBαt + IκBε | Xε | [b] | 0.5 | [3] | 0.3304 |
| 0 → IκBαt | Yα | [c] | 0.00000005 | [3] | 0.000000084 |
| 0 → IκBβt | Yβ | [c] | 0.000000005 | [3] | 0.00000000414 |
| 0 → IκBεt | Yε | [c] | 0.000000005 | [3] | 0.00000000508 |
| IκBαt → 0 | Zα | [b] | 0.0004 | [1] | 0.0003375 |
| IκBβt → 0 | Zβ | [b] | 0.0004 | [3] | 0.0002031 |
| IκBεt → 0 | Zε | [b] | 0.0004 | [3] | 0.0004742 |

**Figure Captions**

Figure 1: Biochemical network model for IKK-IκB-NF-κB-A20 signaling module. Top panel: A schematic description of our comprehensive network model of NF-κB signaling. The arrows indicate activation and the perpendicular lines denote inhibition. In the bottom panel, the comprehensive network model consists of IKK (IκB kinase), IκB isoforms (IκB$i$, $i$=α, β, ε), and A20. NF-κBn and IκB$i$n denote their nuclear components. Squares are for proteins and hexagons are for mRNA. Black arrows indicate either association or dissociation or degradation of proteins while red (blue) arrows denote mRNA (protein) synthesis.

Figure 2: Individual time-series curves and their ensemble average of key protein concentrations are obtained from the computer-generated ensemble of 1000 replicates of a wild type NF-κB signaling system. We compare the computational results with the population-level experimental data from Ref. (Hoffmann et al 2002) side by side. Top left panel: nuclear concentration of NF-κB. The other remaining panels: cytoplasmic concentration of IκBα, IκBβ, and IκBε proteins.

Figure 3: Individual time-series curves and their ensemble average of key protein concentrations are obtained from the computer-generated ensemble consisting of 1000 replicates of a IκB double gene knocked-out mutant. Computational simulation results (left column) are compared with the population-level experimental data (right column) from Ref. (Hoffmann et al. 2002). Top panel: IκBβ and IκBε knocked-out mutant. Mid panel: IκBα and IκBβ knocked-out mutant. Bottom panel: IκBα and IκBε knocked-out mutant.

Figure 4: Individual time-series curves and their ensemble average of key protein concentrations are obtained from the computer-generated ensemble consisting of 1000 replicates of a wild type (right column) and a A20 knocked-out mutant (left column). Computational simulation results are compared with the population-level experimental data from Ref. (Lee et al 2000) side by side. Top panels: nuclear concentration of NF-κB.

Bottom panel: IKK concentration. Other key biochemical species profiles are presented in supporting figure 1.

Figure 5: Dependence of the individual time-series curves and the statistical ensemble average of nuclear NF-κB profiles obtained from IκBβ and IκBε genes knocked-out mutant on the heterogeneity measure χ (i.e., the interval size of the uniform distribution). χ=10 % for top left panel; χ=30 % for top right panel; χ=50 % for bottom left panel; χ=70 % for bottom panel.

Figure 6: Distributions of six dynamic features of nuclear NF-κB profiles are obtained from the ensemble of 1,000 replicates of a wild type (black), A20 knocked-out mutant (red), and three IκB double genes knocked-out mutants (blue, yellow, and green). The six dynamic features are the amplitude of the first peak (First Maximum), the timing of the first peak (First Translocation Time), the First Period, the First Minimum, the Second Maximum, and the Steady State: The last three values are normalized by the First Maximum.

Figure 7: Distributions of the dynamic patterns of the individual time-series curves of nuclear NF-κB profiles are obtained from the ensemble of 1,000 replicates of a wild type, A20 knocked-out mutant, and three IκB double genes knocked-out mutants. Top panel demonstrates four dynamic patterns: (A) single-peaked pattern, (B) under-damped oscillation, (C) hyperbolic pattern, and (D) sustained oscillation. Each individual time-series curve is classified into one of four dynamic patterns.

Figure 8: Individual time-series curves and their ensemble average of the key protein concentrations are obtained from the computer-generated ensemble consisting of 1000 replicates of a wild type system upon stimulation by small dosage (left column) or large dosage (right column).

Figure 9: Distributions of six dynamic features of nuclear NF-κB profiles are obtained from the ensemble of 1,000 replicates of a wild type NF-κB signaling system with a small (red color; TR=0.01) or large (black-color; TR=1) dosage stimulation.

Figure 10: Distribution of the dynamic patterns of nuclear NF-κB profiles from the ensemble of 1,000 replicates of a wild type NF-κB signaling system upon a small (TR=0.01) or a large (TR=1) dosage stimulation. Yellow denotes sustained oscillation; red for damped oscillation; blue for single-peaked pattern.

Figure 11: The individual dose-response curves and their statistical ensemble average from the ensemble consisting of 50 computer-generated replicates of a wild type NF-κB signaling system.

Supporting Figure 1: Individual time-series curves and their ensemble average of key protein concentrations are obtained from the computer-generated ensemble consisting of 1000 replicates of a wild type (right column) and a A20 knocked-out mutant (left column). Computational simulation results are compared with the population-level experimental data from Ref. (Lee et al 2000) side by side. Top panels: cytoplasmic concentration of IκBα protein. Mid panel: concentration of IκBα mRNA. Bottom panel: A20 protein concentration.

Figure 1

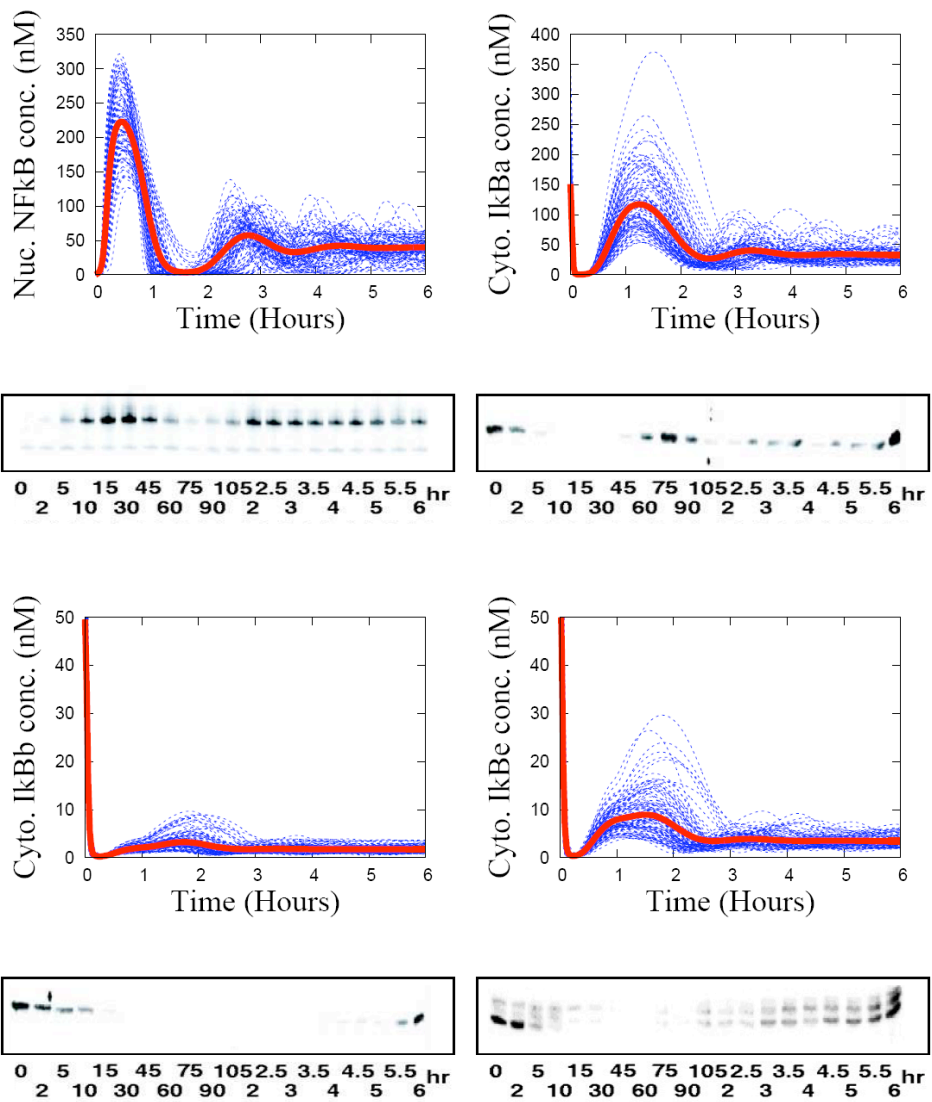

Figure 2

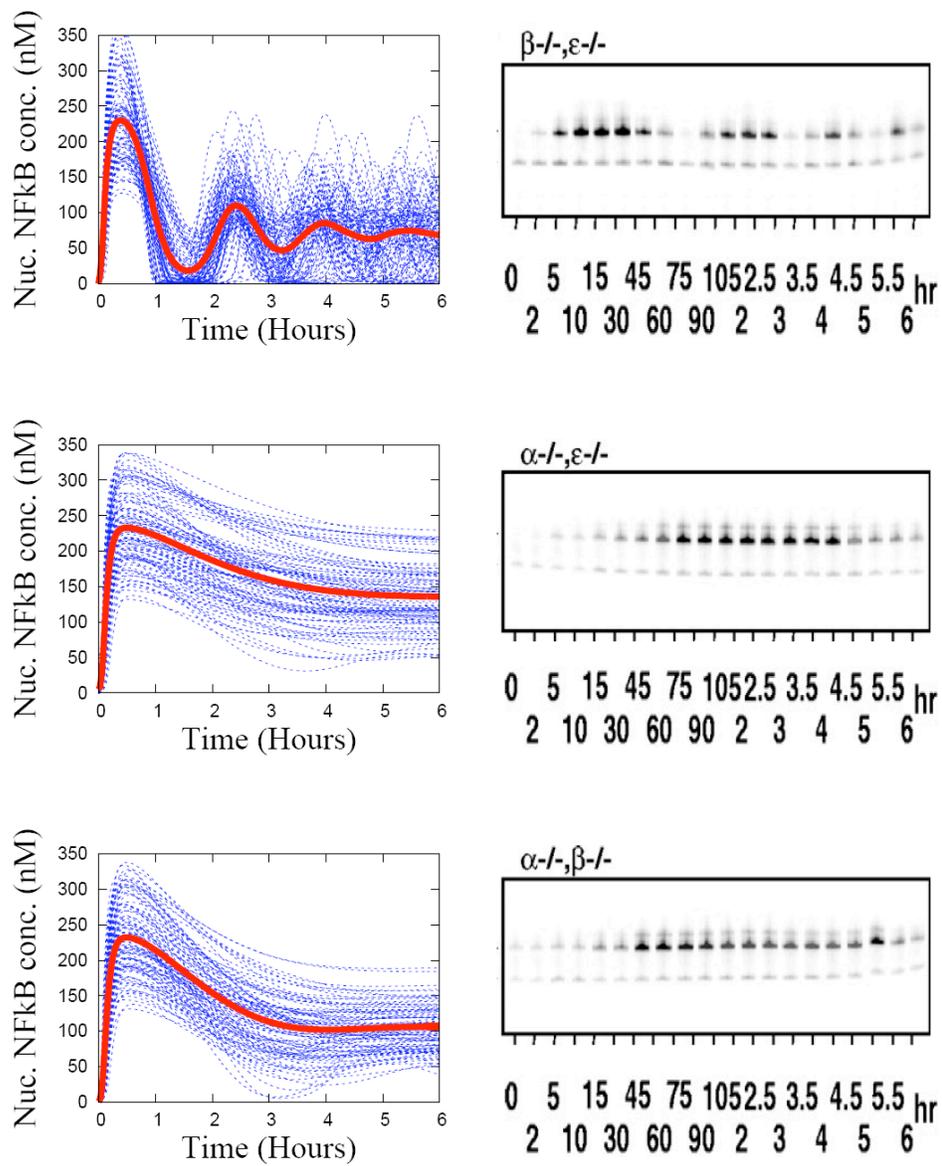

Figure 3

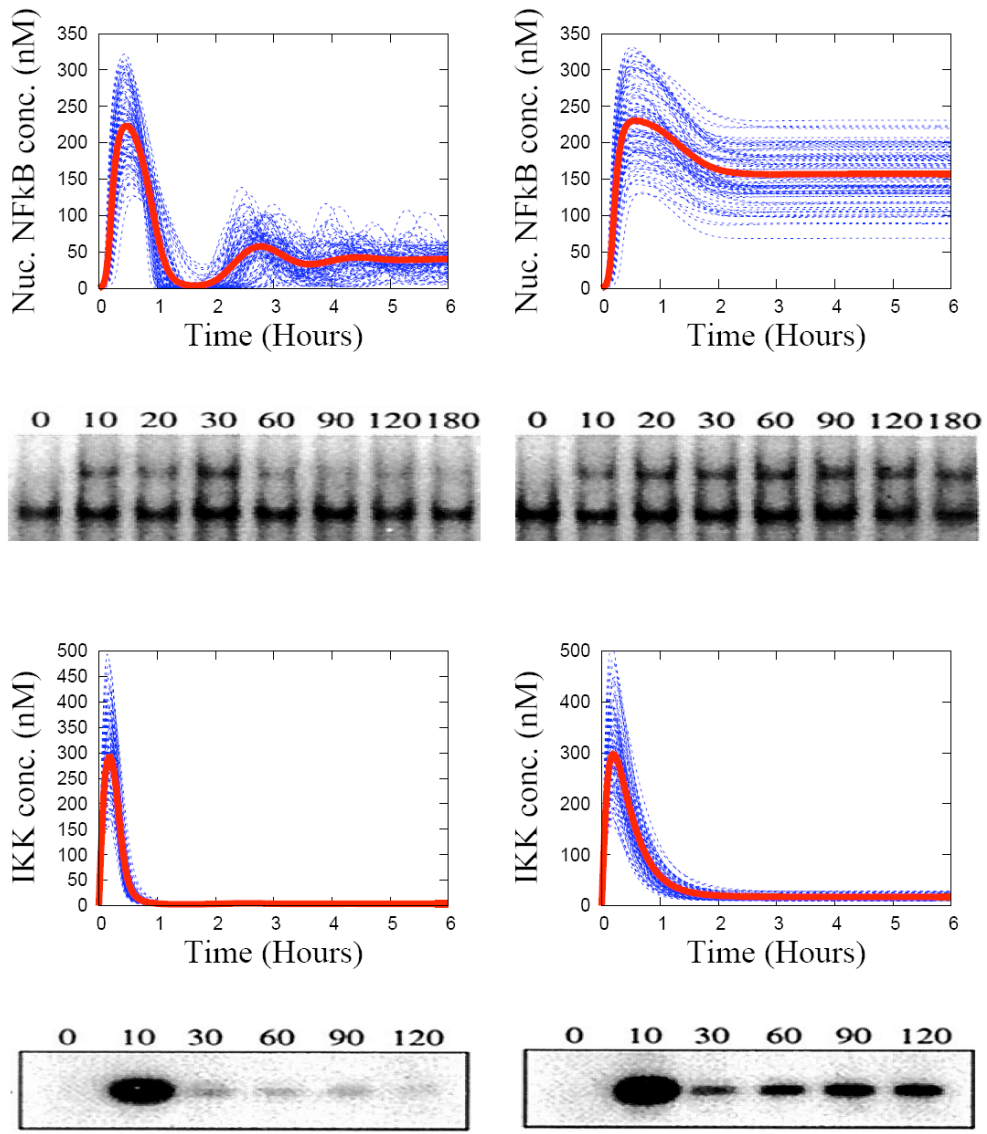

Figure 4

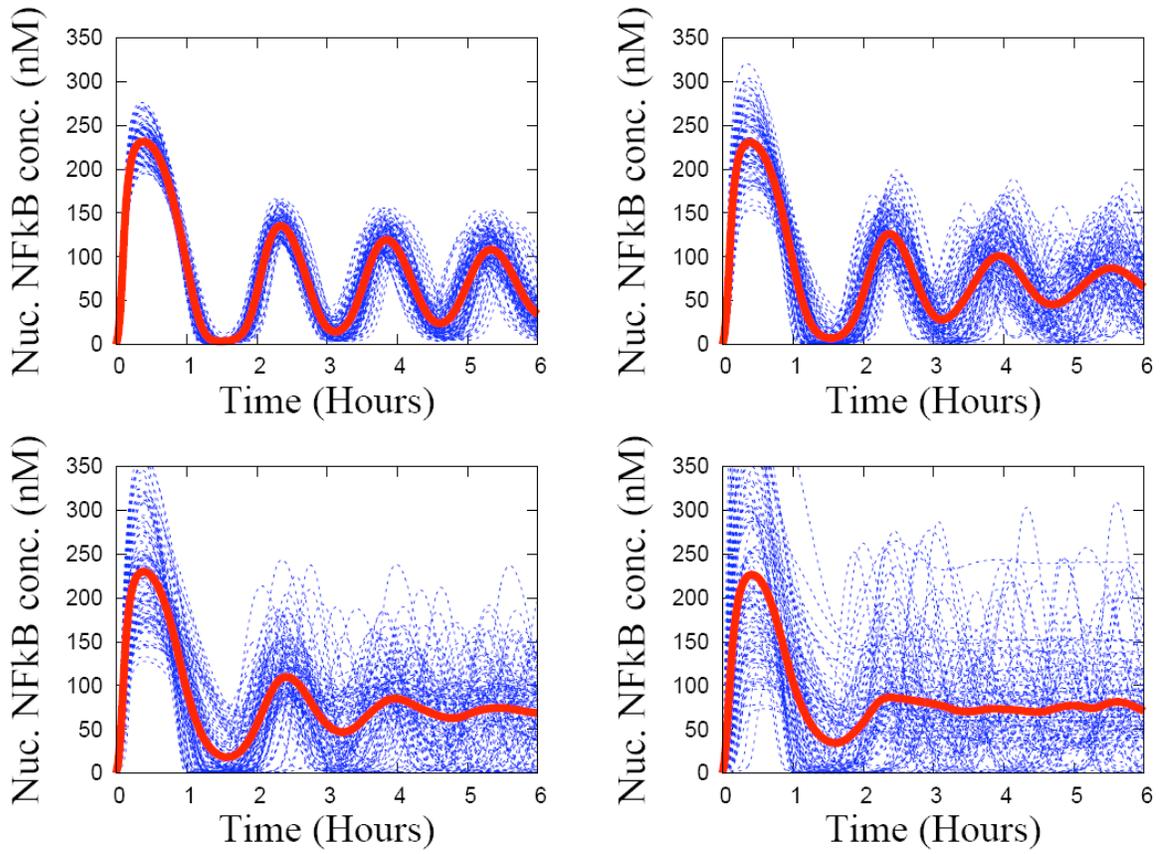

Figure 5

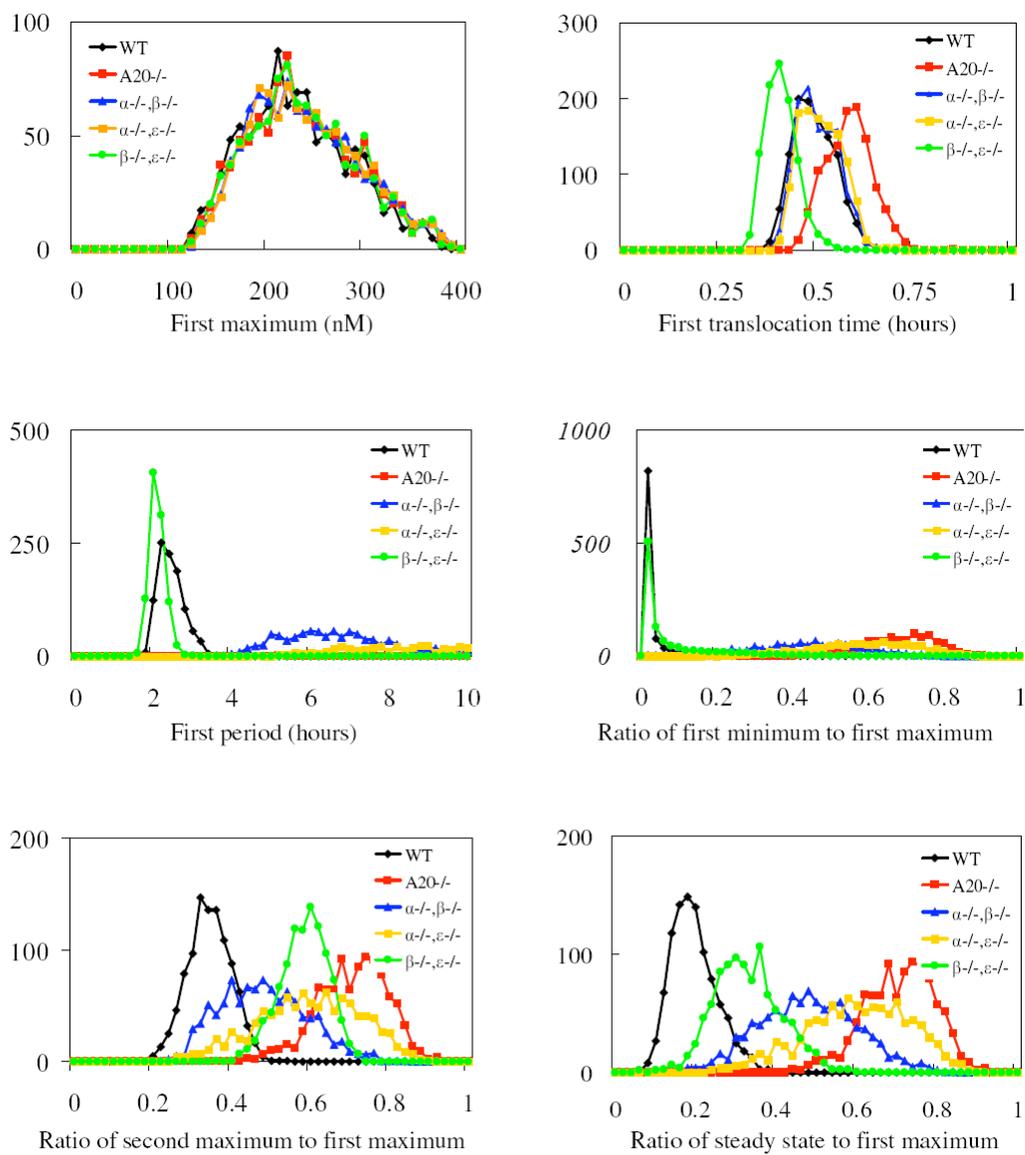

Figure 6

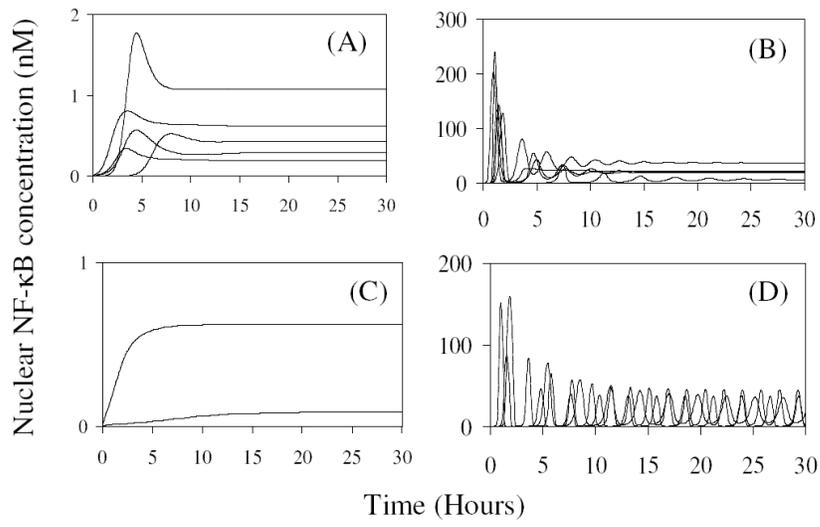

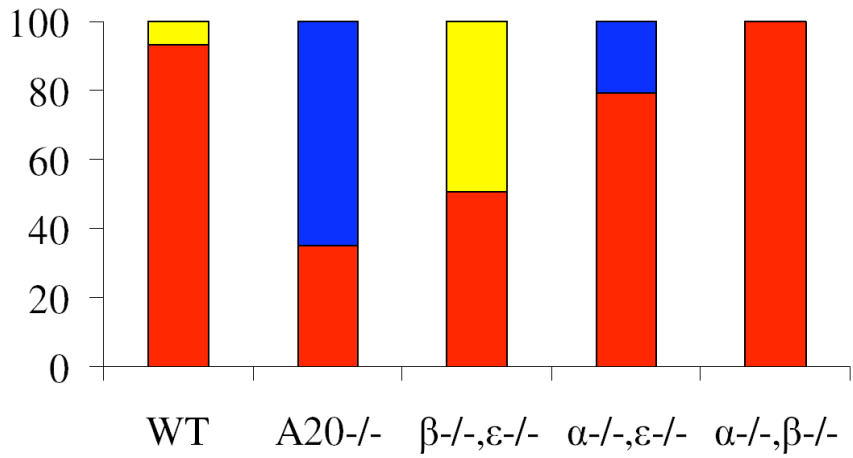

Figure 7

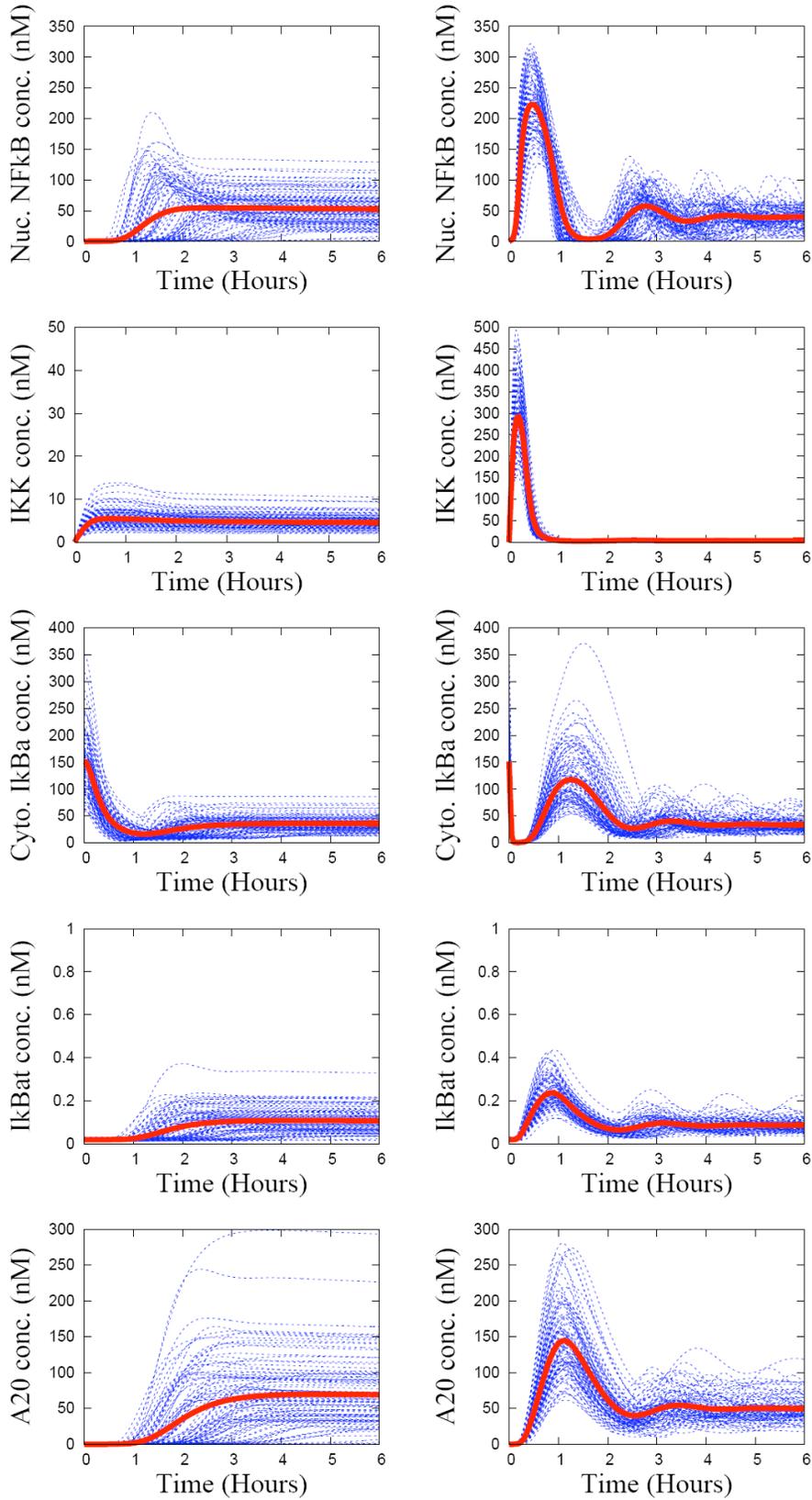

Figure 8

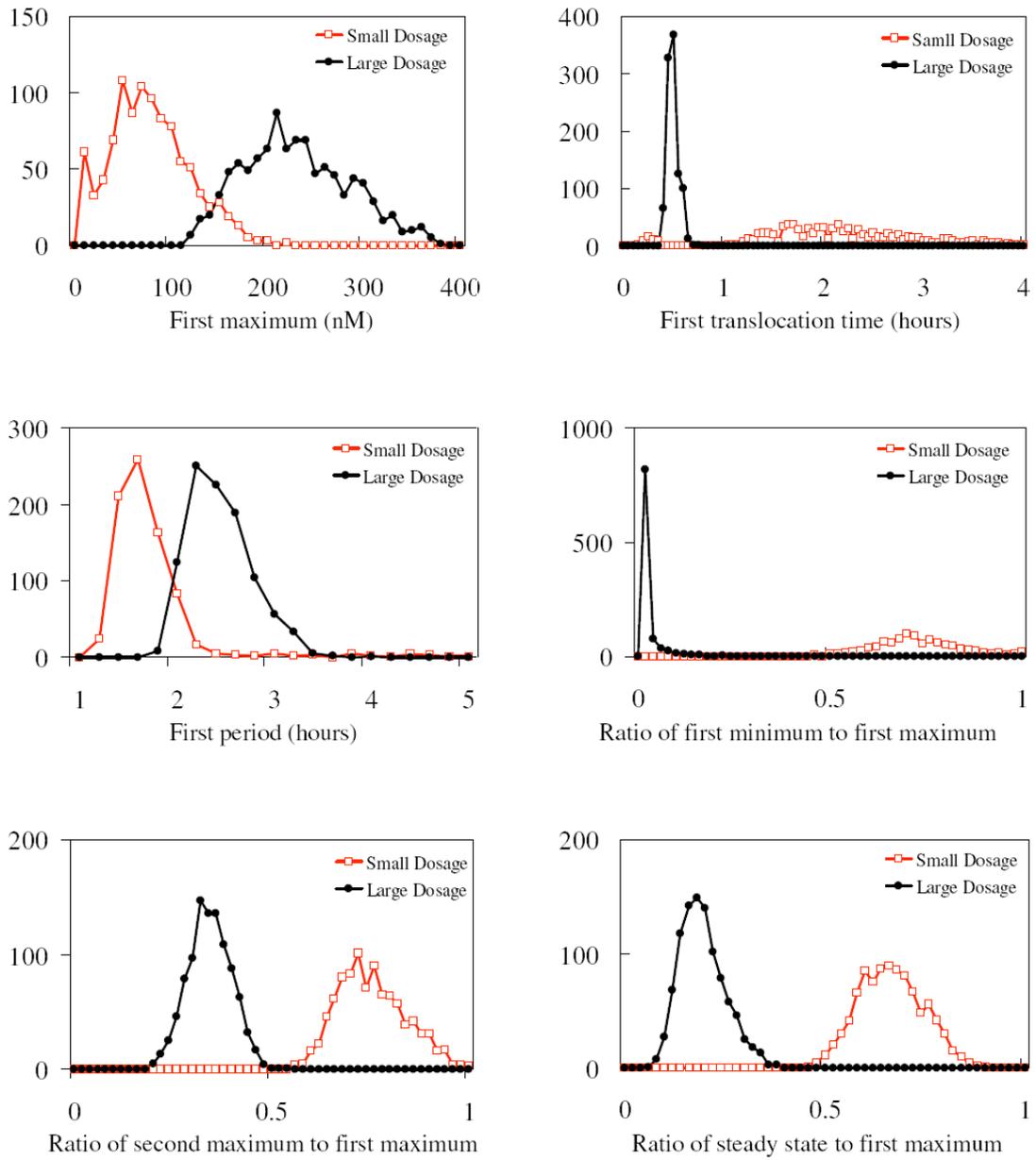

Figure 9

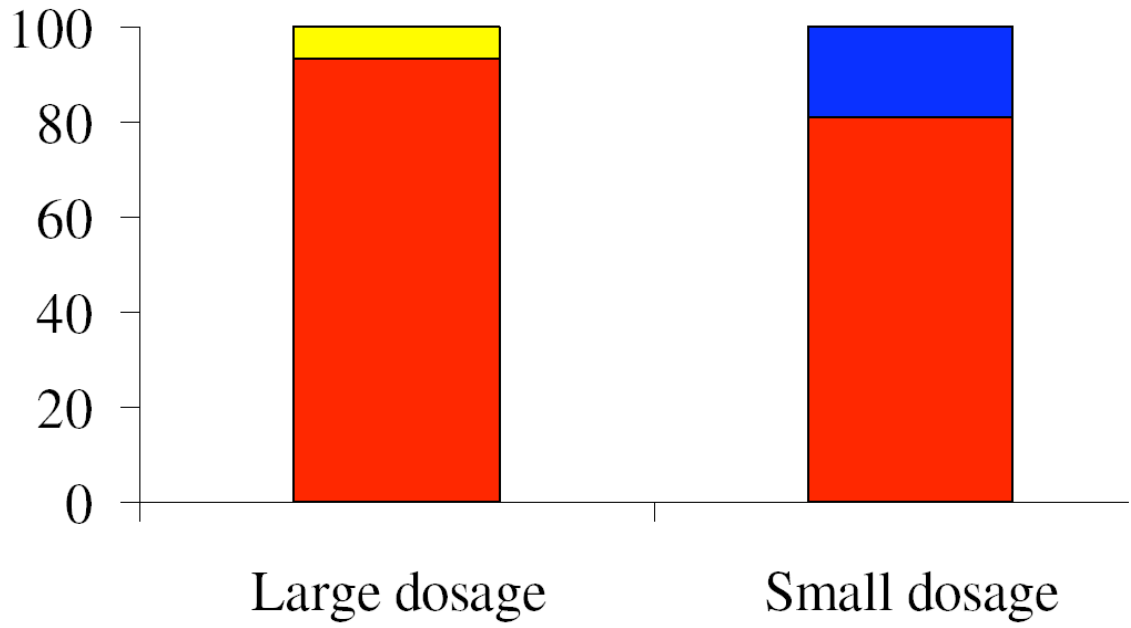

Figure 10

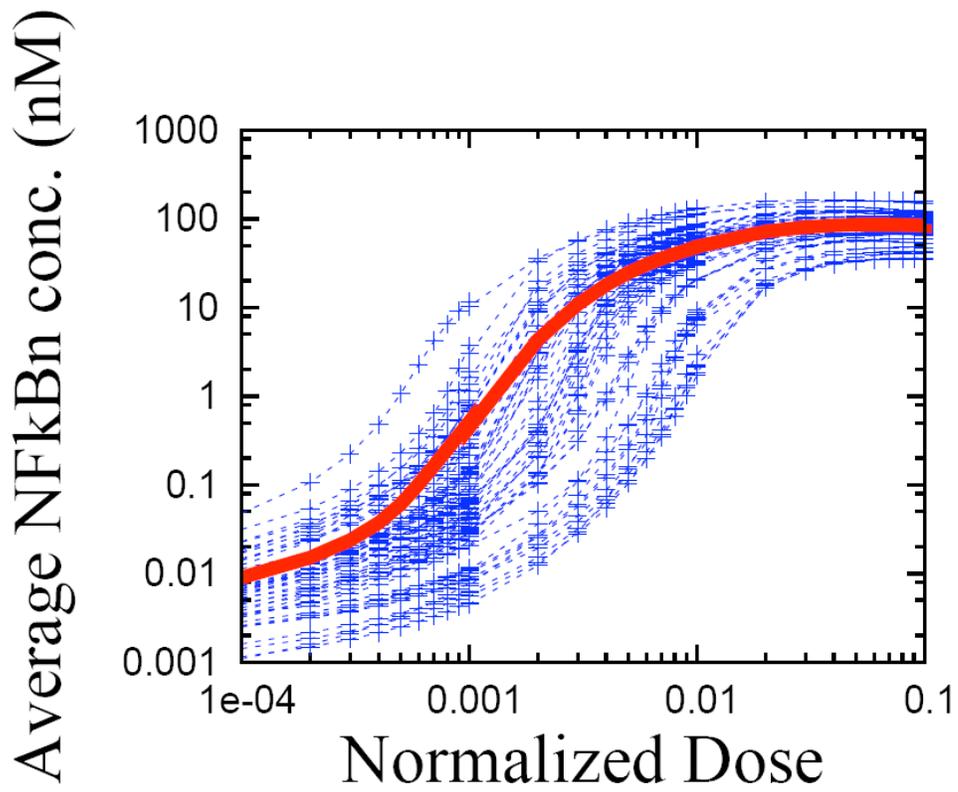

Figure 11

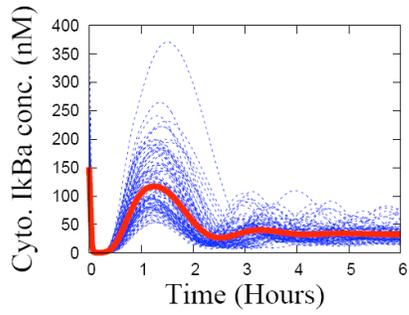
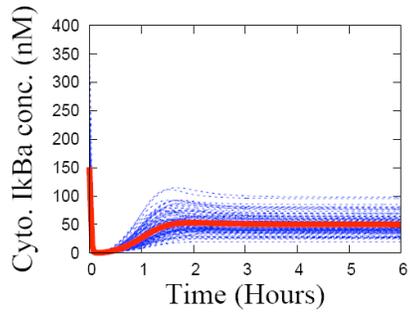

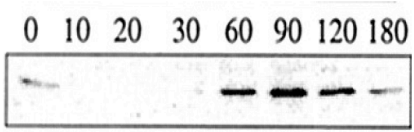
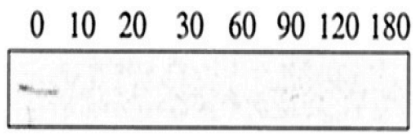

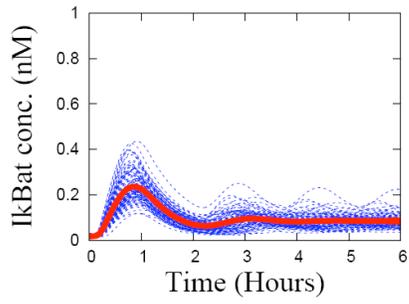
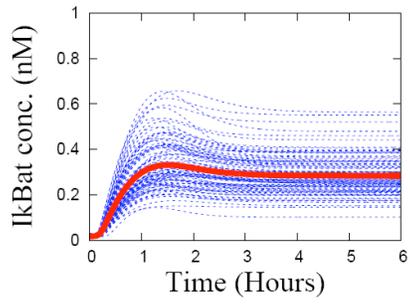

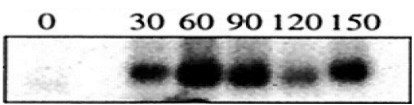
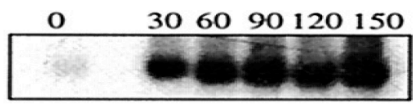

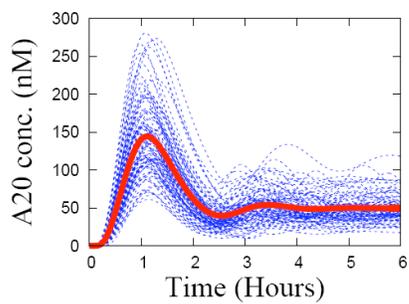

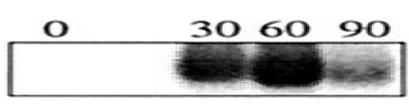

Supporting Figure 1